\documentclass[aps,twocolumn,showpacs,superscriptaddress,amsmath]{revtex4}
\usepackage{graphicx}
\newcommand{\bea}{\begin{eqnarray}}
\newcommand{\beq}{\begin{equation}}
\newcommand{\eea}{\end{eqnarray}}
\newcommand{\eeq}{\end{equation}}
\begin{document}
\title{Spectral spacing correlations 
of nucleus-nucleus collisions at high energies}
\vspace{0.5cm}
\author{R.\ G.\ Nazmitdinov}
\affiliation{Departament de F{\'\i}sica,
Universitat de les Illes Balears, E-07122 Palma de Mallorca, Spain}
\affiliation{Bogoliubov Laboratory of Theoretical Physics,
Joint Institute for Nuclear Research, 141980 Dubna, Russia}
\author{E.\ I.\ Shahaliev}
\affiliation{High Energy Physics Laboratory, Joint Institute for
Nuclear Research, 141980, Dubna, Russia} \affiliation{Institute of
Radiation Problems, 370143, Baku, Azerbaijan}
\author{M.\ K.\ Suleymanov}
\affiliation{High Energy Physics Laboratory, Joint Institute for
Nuclear Research, 141980, Dubna, Russia}
\author{S.\ Tomsovic \footnote{permanent address: Department of Physics 
and Astronomy, Washington State University, Pullman, WA  99164-2814}}
\affiliation{Max-Planck-Institut f\"ur Physik komplexer Systeme, N\"othnitzer 
Stra$\beta$e 38, D-01187 Dresden, Germany}

\date{\today}

\begin{abstract}
We propose a novel approach to the analysis of experimental data obtained
in relativistic nucleus-nucleus collisions which borrows from methods developed 
within the context of Random Matrix Theory.  It is applied to the detection of 
correlations in momentum distributions of emitted particles.  We find good agreement 
between the results obtained in this way and a standard analysis based on the 
two-pair correlation function often used in high energy physics.
The method introduced here is free from unwanted background contributions.
\end{abstract}
\maketitle

There is currently an enormous effort underway to detect signals of possible 
transitions between different phases of a composite system produced  in high 
energy nucleus-nucleus collisions 
(cf \cite{1,2}).  It is anticipated that in central collisions, at energies that 
are and will be soon available
at SPS(CERN), RHIC(BNL) and LHC(CERN), the nuclear density may exceed the density 
of stable nuclei by an order of magnitude.  Under such extreme conditions, according 
to a generally held beliefs, the final product of a heavy ion collision would be 
a composite system that consists of free nucleons, quarks and a quark-gluon plasma.   
Various methods have been proposed to identify possible 
manifestations of such a quark-gluon plasma.  Often though, results based on such 
methods are sensitive to assumptions made concerning the background measurements 
and mechanisms included in the corresponding model.  In addition, the identification 
of the quark-gluon plasma is made more difficult due to a large multiplicity of 
secondary particles created at these collisions.   The formulation of a reliable 
criterion for the selection of meaningful signals is, indeed, an important objective 
in relativistic heavy ion collisions physics. 

In a preliminary report \cite{sh1}, we suggested that tools from Random Matrix 
Theory (RMT) \cite{meh} might be useful in illuminating the presence of correlations 
in the spectral (momentum) distribution of secondary particles produced in 
nucleus-nucleus collisions at high energy.  The RMT approach does not depend on 
the background of measurements and relies only on fundamental symmetries preserved 
in heavy-ion collisions.  Furthermore, the larger the multiplicity, the better 
the applicability of the RMT tools, and thus, its predictive power.  In the 
present paper, we demonstrate that  the RMT analysis is very sensitive to 
spectral spacing correlations present in the nucleus-nucleus collision data, more 
so than the standard tools used for such an analysis. 

Here, we make use of the experimental data that have been obtained with the 2-m propane 
bubble chamber of LHE, JINR~\cite{[8],[9]}. The chamber, placed in a magnetic field 
of 1.5 T, was exposed to beams of light relativistic nuclei at the Dubna Synchrophasotron.  
Nearly all secondary particles, emitted at a 4$\pi$ total solid angle, were detected 
in the chamber.  All negative particles, except those identified as electrons, are 
considered as $\pi^-$-mesons. The contamination from misidentified electrons and 
negative strange particles does not exceed 5$\%$ and 1$\%$, respectively. 
The average minimum momentum for pion registration is about 70 MeV/c. The protons 
were selected by a statistical method applied to all positive particles with a 
momentum of $|p|>500$ MeV/c (we identified slow protons with $|p|\le 700$ MeV/c 
by ionization in the chamber). In this experiment, there are 37792 ${}^{12}{CC}$  
interaction events at a momentum of 4.2A GeV/c (for greater discussion of the 
details see~\cite{[9]}) containing 7740 events with more than ten tracks of charged 
particles.

The basis of our approach derives from RMT \cite{meh}, which was originally introduced 
to explain the statistical fluctuations of neutron resonances in compound 
nuclei \cite{P65} (see also Ref.\ \onlinecite{Brody}).  There the precise 
heavy-compound-nuclear Hamiltonian is unknown or rather poorly known, and there 
is a large number of strongly interacting degrees of freedom.  Wigner first suggested 
replacing it by an ensemble of Hamiltonians which describe all possible 
interactions \cite{Wigner}.  The theory assumes that the Hamiltonian belongs to 
an ensemble of random matrices that are consistent with the fundamental symmetries 
of the system.  In particular, since the nuclear interaction preserves time-reversal 
symmetry, the relevant ensemble is the Gaussian Orthogonal Ensemble (GOE).  Whereas 
if time-reversal symmetry were broken, the Gaussian Unitary Ensemble (GUE) would 
be the relevant ensemble. The GOE and GUE correspond to ensembles of real symmetric  
matrices and of complex Hermitian matrices, respectively.  Besides the general symmetry 
considerations, no other property of the system under consideration is taken into 
account.

If one supposes that the momenta distributions of secondary particles produced in 
nucleus-nucleus collisions may be treated in analogy with eigenstates of a 
composite system, just like the eigenstates of the compound nucleus, then the 
statistical analysis methods introduced by Dyson and Mehta can be applied to 
the LHE collision data \cite{P65}.  The difference between energy and momentum 
is not essential for pions, and we assume that the proton mass does not significantly 
affect the correlation function.  
Note also, that here we are dealing with the momentum 
distribution in the target rest frame only, postponing its comparison to that in 
the center of mass frame, which is more natural for description of interaction.
Based on this supposition, the ordered sequence 
of ``energy levels'' $\{E_i\}, i=1,...,N$ comes from the momentum distribution and 
it has an average density of states denoted $\rho_{\it av}(E)$.  From this sequence 
a new one is obtained  by the unfolding procedure 
of the original spectrum $\{E_i\}$ through the mapping $E \rightarrow x$
\beq
\label{m}
x_i=\int_0^{E_i}\rho_{\it av}(E^\prime)dE^\prime=\int_0^{x_i}dx^\prime=\zeta(E_i), \qquad i=1,...,N
\eeq
Here, $\zeta(E)$ is the smooth function giving the mean number of eigenvalues less 
than or equal to $E$ of the exact eigenvalue distribution $N(E)$, which is often 
referred to as the staircase function due to its appearance (see Ref.\onlinecite{sh1}).  
The smooth part $\zeta(E)$ can be determined either from semiclassical arguments 
or by using a polynomial/spline interpolation for the exact staircase function.  
The momentum distribution data (see Fig.1 in Ref.\onlinecite{sh1}) were approximated 
by a polynomial function of sixth order and the distributions of various 
spacings $s_i$ from the 7740 events satisfiy the condition of $\chi^2$ per degree 
of freedom is less than unity.

\begin{figure}[ht]
\includegraphics[height=0.25\textheight,clip]{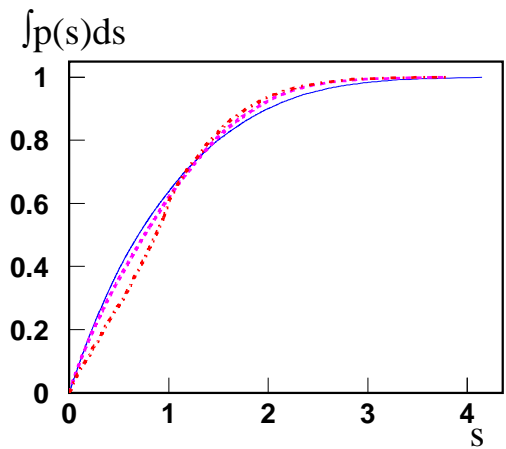}
\caption{(Color online) Cumulative  spacing momentum distributions 
            for different  regions of measured momenta:
        a)$0.1<|p|<1.14$ GeV/c (solid line); b)$1.14<|p|<4.0$GeV/c (dashed line);
        c)$4.0<|p|<7.5$ GeV/c (dot-dashed line).
        The solid line exhibits  the Poisson distribution, while the
        dot-dashed line is close to the  Wigner-Dyson type distribution.} 
\label{fig1}
\end{figure}
The effect of the mapping is that the sequence $\{x_i\}$ has on average a
constant unit mean spacing (a constant unit density), irrespective of the
particular form of the function $\zeta(E)$ \cite{boh2}.  The spacings are defined as
$s_i=x_{i+1}-x_i$ between two adjacent points and they are 
collected in a histogram, which gives the probability density.  If the "events" $\{x_i\}$ 
are independent, then the form of the histogram must follow $p(s)=exp(-s)$ known 
as the Poisson density.  On the other hand, if the levels are repelled 
(anticorrelated) as in the GOE, the density is approximately given by the Wigner 
surmise form $p(s)=\frac{\pi}{2}s \exp(-\frac{\pi}{4}s^2)$.  Interestingly,  
the spacing probability density approximately follows the Wigner surmise 
for high energies, whereas at relatively
low energies the corresponding spacing density is maximum at the origin and 
nearly the Poisson density \cite{sh1}.  In an eigenspectrum, the Poisson density 
arises where there is a dominance of many crossings between different eigenenergies, 
whereas the Wigner surmise reflects the
tendency to avoid clustering of levels.  In turn, the crossings are usually observed
where there is no mixing between states that are characterized by different good
quantum numbers, and the anticrossings signal a strong mixing due to
a perturbation brought about by either external or internal sources 
(cf Refs.\onlinecite{P65,Brody}). 

The transition from one probability density to the other has been used, for example, 
in nuclear structure physics to study the stabilization of octupole deformed shapes 
and transition from the chaotic to the 
regular pattern in the classical limit \cite{H94}.   Therefore, such an analysis 
can provide the first hint of some structural changes at different parameters of 
the system under consideration, in particular, in 
different energy (momentum) intervals. Figure \ref{fig1} shows the integrated 
momentum spacing density for experimental data, demonstrating a gradual transition 
from a Poisson-like density toward a Wigner-like density with the increase of 
the absolute value of the momentum distributions.

In order to elucidate the degree of correlations for a stationary spectrum with 
unit average spacing
Dyson introduced the {\it k-level correlation functions}
\begin{eqnarray}
&R_k(x_1,...,x_k)=\frac{N!}{(N-k)!}\int ...\int P_k(x_1,...,x_N)dx_{k+1}...dx_N\nonumber\\ 
&1\leq k\leq N,
\label{p1}
\end{eqnarray}
where $P_k(x_1,...,x_N)dx_{1}...dx_N$ gives the probability of having one eigenvalue 
at $x_1$, another at $x_2$..., another at $x_N$ each within the interval 
$\{x_i, x_i+dx_i\}$. By integrating $P_k(x_1,...,x_N)$ over all variables but one, 
in the
limit $N\rightarrow\infty$, one obtains the ensemble averaged density
\begin{equation}
\tilde{\rho}(x_1)=\int ...\int P_k(x_1,...,x_N)dx_{2}...dx_N
\end{equation}
 which is normalised to unity.  From Eq.(\ref{p1}) it follows that 
 $R_1(x_1)=N\tilde{\rho}(x_1)$ 
and $R_k(x_1,...,x_k)dx_1....dx_k$ is the probability, irrespective  of 
the index, of finding  one level within of each of the intervals $[x_i,x_i+ds]$.  
From the above definition it follows that $R_1(x)=1$.  With the aid of the 
definition (\ref{p1}), by integrating $R_{k+1}$ one obtains
\begin{equation}
\int R_{k+1}(x_1,...,x_{k+1})dx_{k+1}=(N-k)R_{k}(x_1,...,x_{k})
\end{equation}
\begin{figure}[ht]
\includegraphics[height=0.25\textheight,clip]{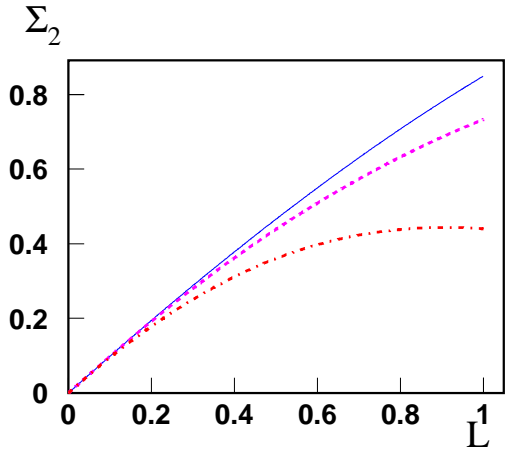}
\caption{(Color online) 
Number variance  $\Sigma^2(L)$ for three
different region of the experimental spacing momentum distributions:
        a)$0.1<|p|<1.14$ GeV/c (solid line); b)$1.14<|p|<4.0$GeV/c (dashed line);
        c)$4.0<|p|<7.5$ GeV/c (dot-dashed line).} 
\label{fig2}
\end{figure}
It is difficult to work directly with the $R_k$ functions.  One important and 
more convenient measure of correlation that was introduced is based on the number 
statistic $n(L)$ which is defined to be
the number of levels in an energy interval of length $L$. Since the spectrum was 
unfolded,
the average number statistic $\langle n(L) \rangle =L$ is independent of the spectrum.
However, the variance of $n(L)$
\begin{equation}
\Sigma^2(L)=\langle[n(L)-\langle n(L)\rangle]^2\rangle
\label{sig}
\end{equation}
does depend on the spectrum considered. For the Poisson density (see \cite{meh})
\begin{equation}
\Sigma^2(L)=L\quad ,
\label{sigP}
\end{equation}
and for the GOE, the exact asymptotic expression is
\begin{eqnarray}
&&\Sigma^2(L)=\frac{2}{\pi^2}\Bigg{[}ln(2\pi L)+\gamma+1+\frac{1}{2}[Si(\pi L)]^2
-\frac{\pi}{2}Si(\pi L)\nonumber\\
&&-\cos(2\pi L)-Ci(2\pi L)+
\pi^2L\left[1-\frac{2}{\pi}Si(2\pi L)\right]\Bigg{]}
\end{eqnarray}
Here $\gamma$ is the Euler constant and $Si$, $Ci$ are the sine and cosine integrals,
respectively.
The number variance $\Sigma^2(L)$ calculated using the optimal implemention of 
the definition in Eq.(\ref{sig}) is shown in Fig.\ \ref{fig2}. Indeed, this 
quantity manifests the Poisson
statistics (Eq.\ (\ref{sigP})) for experimental spectra with a low momenta distribution. 
On the other hand, one again observes a clear indication on 
the presence of correlations for large momenta.
\begin{figure}[hb]
\includegraphics[height=0.25\textheight,clip]{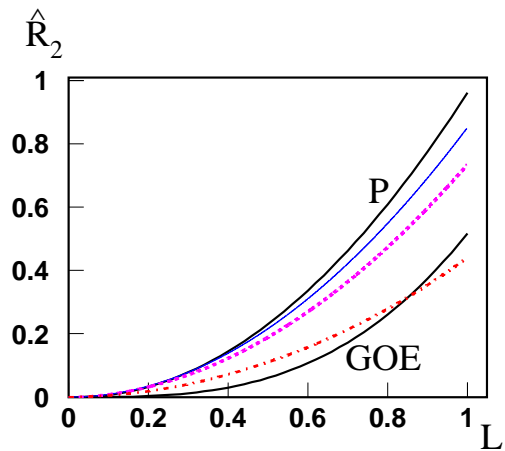}
\caption{(Color online) Two-point  correlation function ${\hat R_2}(s=L)$
                for different  regions of measured momenta:
        a)$0.1<|p|<1.14$ GeV/c (solid line); b)$1.14<|p|<4.0$GeV/c (dashed line);
        c)$4.0<|p|<7.5$ GeV/c (dot-dashed line).
        The solid lines, denoted as P and GOE, display the characteristic  
	limits for Poisson and GOE ensembles, respectively.} 
\label{fig3}
\end{figure}

For the analysis of fluctuations, it is more convenient to use pure
k-point functions \cite{BHP}
\begin{equation}
\hat{R}_k(L)=\int_0^L...\int_0^L R_{k}(x_1,...,x_{k})dx_1...dx_{k}
\end{equation}
The function $\hat{R}_k(L)/k!$ gives the probability that 
an interval of length $L$ (for small $L$) contains $k$ levels.
In RMT most emphasis has been put on the two-point 
correlation function ${\hat R_2}(x_1,x_2)$ or density-density correlation function.
The two-point correlation function is the probability density to find two eigenvalues 
$x_i$ and $x_j$ at two given energies, irrespective of the position of all
other eigenvalues. 
The function $R_2(x_1,x_2)$ depends only on the relative variables $s=x_1-x_2$.
The variance $\Sigma^2$ is connected to $\hat R_2$ through the following relation
\begin{eqnarray}
\hat R_2(L)=&&\int_0^L\int_0^L R_{2}(x_1,x_{2})dx_1dx_{2}-2\int_0^L(L-s)R_2(s)ds\nonumber\\
=&&\Sigma^2(L)+L(L-1)
\label{r2}
\end{eqnarray}
The two-level correlation function $R_2(x_1,x_2)$  determines
the basic fluctuation measures related to Wigner's level repulsion and
the Dyson-Mehta {\it long-range order}, i.e., large correlations between distant 
levels. Bohigas {\it et al} (1985) \cite{BHP} provided a thorough analysis 
of level repulsion and long-range correlations (rigidity) for different correlation 
functions. To understand the distinct role played by level repulsion and long-range order 
in the momentum density, we compare our numerical results with 
analytical expressions from Table 1 of Ref.\onlinecite{BHP} for the Poisson 
ensembles (there is neither level repulsion nor long-range order) and for the
GOE case (this ensemble exhibits both features).  The two-point correlation 
function ${\hat R_2}(x_1,x_2)$ calculated from Eq.(\ref{r2}) is shown in Fig.\ref{fig3}.  
Even though there are small deviations from Poisson $(\hat R_2(L)= L^2)$ and 
GOE $(\hat R_2(L)= \pi^2L^3/18)$ predictions, the experimental results for the 
momentum distributions reproduce surprisingly well both limits. 

The validity of the RMT analysis is confirmed by an independent analysis of 
the data with the aid of the standard pair-correlation function 
(see, for example, Ref.\onlinecite{gol} and references therein):
\begin{equation}
R(y_1,y_2)=\sigma \frac{d^2\sigma/dy_1dy_2}{(d\sigma/dy_1)(d\sigma/dy_2)}-1
\end{equation}
Here, the quantity $\sigma$ is the cross section of the inclusive reaction and
$y=\frac{1}{2}ln\frac{E+P_{||}}{E-P_{||}}$ is the rapidity, which depends on the
particle energy $E$ and its longitudinal momentum $P_{||}$.
The rapidity is one of the main characteristics widely used in relativistic
nuclear physics (see \cite{b2,b4}).
\begin{figure}[ht]
\includegraphics[height=0.38\textheight,clip]{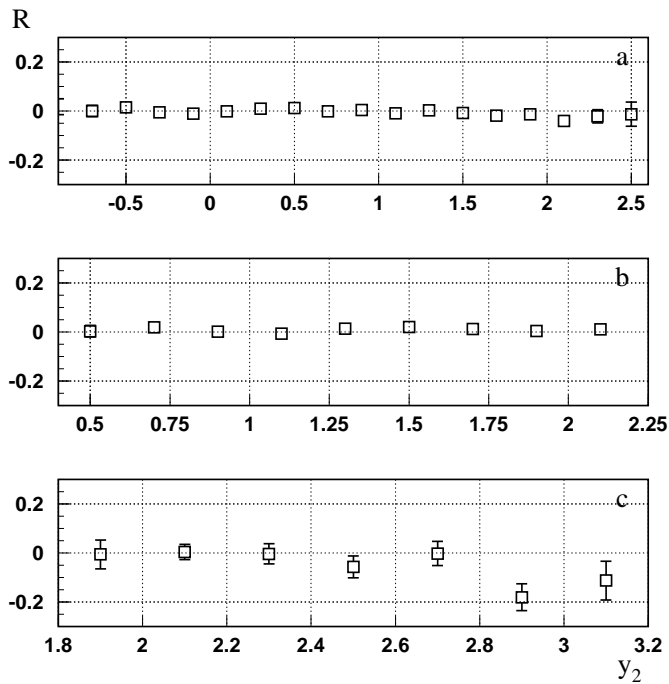}
\caption{Integrated two-pair correlation functions for 
        particles obtained in CC-interactions (see the text) 
                for different  regions of measured momenta:
        a)$0.1<|p|<1.14$ GeV/c; b)$1.14<|p|<4.0$GeV/c;
        c)$4.0<|p|<7.5$ GeV/c.
	} 
\label{fig4}
\end{figure}
The pair-correlation function manifests the difference between probability density 
of two-particle events and the product of the probability densities of independent 
particle events. It vanishes if the particle rapidities are independent.
Figure \ref{fig4} demonstrates the results for particles 
obtained in CC-interactions.  For the sake of illustration, we integrate the 
function $R(y_1,y_2)$ over one of the variables, say $y_1$, and consider the 
dependence on $y_2$. For different momentum distributions, there are three intervals 
of integration for the variable $y_1$:
a)for $0.1<|p|<1.14$ GeV/c the function $R(y_1,y_2)$ is integrated in the 
interval $-0.9<y_1<2.5$;
b)for $1.14<|p|<4.0$GeV/c  it is integrated in the interval $0.5<y_1<2.4$; and
c)for $4.0<|p|<7.5$ GeV/c it is integrated in the interval $2.5<y_1<3.5$.  
There are, of course, some experimental  errors which are not seen in the behaviour 
of the function $R=\int_{y_1}R(y_1,y_2)dy_1$ on Figs.\ref{fig4}a,b due to the 
large data set.  These errors do not spoil, however,  the results in the region c), 
where experimental data on the multiparticle production exhibits an indication 
of the existence of correlations between particles in the region $4.0<|p|<7.5$ GeV/c.  
The physical origin of the correlations is beyond the scope of the present paper 
and will be discussed elsewhere.
One observes that the predictions based on the standard pair-correlation function 
$R$ are consistent with the predictions based on the RMT analysis.  However, 
the RMT two-point correlation function  magnifies the presence of correlations 
manifested in the standard pair-correlation function (compare Figs.\ref{fig3} and 
\ref{fig4}c).

Summarizing, we propose an analysis of relativistic nuclear collision data based 
on ideas from RMT. 
The approach is free from various assumptions concerning the background of the 
measurements and it provides reliable information about correlations induced by 
external or internal perturbations.  All these features make our proposal a quite 
promising avenue for the future analyses of data produced in heavy ion collision 
experiments.

\section*{Acknowledgments}
We are grateful to  Aleksandr Golokhvastov for the fruitful discussions on 
properties of the two-pair correlation function.  This work is partly supported 
by Grant No. FIS2005-02796 (MEC, Spain) and 
by Grant RNP.2.1.1.5409 of the Ministry of Science and 
Education of the Russian Federation. S.\ T.\ gratefully acknowledges support 
from the US National Science Foundation grant PHY-0555301.
R. G. N. gratefully acknowledges support from the Ram\'on y Cajal programme (Spain).

\end{document}